\documentclass[10pt,a4paper,twocolumn,aps,superscriptaddress,Xlinenumbers]{revtex4-1}
\usepackage[utf8]{inputenc}
\usepackage{color}
\usepackage{amsmath}
\usepackage{amssymb}
\usepackage{graphicx}
\usepackage[unicode=true,pdfusetitle,
 bookmarks=true,bookmarksnumbered=false,bookmarksopen=false,
 breaklinks=false,pdfborder={0 0 0},pdfborderstyle={},backref=false,colorlinks=true]
 {hyperref}
\hypersetup{
 citecolor=blue}

\makeatletter
\pdfpageheight\paperheight
\pdfpagewidth\paperwidth
\graphicspath{{./}{./figs/}} 
\makeatother

\begin{document}
\title{Multistable Kuramoto splay states in a crystal of mode-locked laser
pulses}
\author{T. G. Seidel}
\affiliation{Institute for Theoretical Physics, University of Münster, Wilhelm-Klemm-Str.9
48149 Münster, Germany}
\affiliation{Departament de Física and IAC$^{3}$, Universitat de les Illes Balears,
C/ Valldemossa km 7.5, 07122 Mallorca, Spain}
\author{A. Bartolo}
\affiliation{Institut d'Electronique et des Systèmes, CNRS UMR5214, 34000 Montpellier,
France}
\affiliation{Université Côte d'Azur, CNRS, Institut de Physique de Nice, 06200
Nice, France}
\author{A. Garnache}
\affiliation{Institut d'Electronique et des Systèmes, CNRS UMR5214, 34000 Montpellier,
France}
\author{M. Giudici}
\affiliation{Université Côte d'Azur, CNRS, Institut de Physique de Nice, 06200
Nice, France}
\author{M. Marconi}
\affiliation{Université Côte d'Azur, CNRS, Institut de Physique de Nice, 06200
Nice, France}
\author{S. V. Gurevich}
\affiliation{Institute for Theoretical Physics, University of Münster, Wilhelm-Klemm-Str.9
48149 Münster, Germany}
\affiliation{Departament de Física and IAC$^{3}$, Universitat de les Illes Balears,
C/ Valldemossa km 7.5, 07122 Mallorca, Spain}
\author{J. Javaloyes}
\affiliation{Departament de Física and IAC$^{3}$, Universitat de les Illes Balears,
C/ Valldemossa km 7.5, 07122 Mallorca, Spain}
\begin{abstract}
We demonstrate the existence of a multiplicity of co-existing frequency
combs in a harmonically mode-locked laser that we link to the splay
phases of the Kuramoto model with short range interactions. These
splay states are multistable and the laser may wander between them
under the influence of stochastic forces. Consequently, the many pulses
circulating in the cavity are not necessarily coherent with each other.
We show that this partially disordered state for the phase of the
optical field features regular train of pulses in the field intensity,
a state that we term an incoherent crystal of optical pulses. We provide
evidence that the notion of coherence should be interpreted by comparing
the duration of the measurement time with the Kramers' escape time
of each splay state. Our results are confirmed experimentally by studying
a passively mode-locked vertical external-cavity surface-emitting
laser.
\end{abstract}
\maketitle

The realization of mode-locking (ML) has been a milestone of laser
physics as it allowed to generate the ultrashort pulses that are of
paramount importance in many fields including medicine, metrology
and communications \citep{H-JSTQE-00}. The term ML stems from the
synchronous oscillation, i.e. the phase locking of many electromagnetic
modes in a cavity. The emergence of such macroscopic coherent states,
either spontaneously or under periodic forcing, can be seen as the
critical point of an out-of-equilibrium ferromagnetic phase transition
\citep{GP-PRL-02,WRG-PRL-05,LCF-PRL-09} establishing a link between
modal self-organization and frequency combs in active cavities \citep{HAP-PRL-20,CPP-PRL-21}.
The applications of ML encompass radio-over-fibre \citep{NAW-MTT-95},
two-photon absorption microscopy \citep{K-OL-07} or dual comb spectroscopy
\citep{LMW-SCI-17}. The fundamental importance of mode-locked lasers
is demonstrated by their link with dissipative solitons \citep{GA-NAP-12},
their generalization to spatio-temporal systems~\citep{WYD-PRL-11,WSP-Nature-20,DXL-PRL-22,WRC-Optica-22}
or their capability as Ising Photonic Machines to solve NP hard problems
and perform Boltzmann sampling \citep{TYM-AXV-16,IIH-NAP-16,TTY-QST-17}.

\begin{figure}
\includegraphics[width=1\columnwidth]{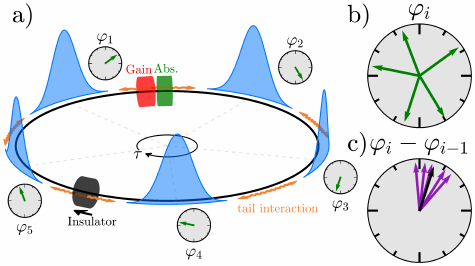}\caption{a) Schematic of a unidirectional ring cavity mode-locked by a saturable
absorber and operated in the HML$_{5}$ regime. Each pulse possesses
a phase $\varphi_{i}$, as indicated by the small clocks. The phases
and their differences are shown in b) and c) while the black arrow
in c) denotes to value of the order parameter $b$ representing the
average coherence.}
\label{fig:setup}
\end{figure}

Soon after their inception, it was demonstrated that a pulsating laser
can also operate in the so-called harmonic mode-locked (HML) regime,
a state in which the laser cavity supports a train of $N\in\mathbb{N}$
equidistant pulses (denoted here as HML$_{N}$) \citep{BKS-JQE-72},
see Fig.~\ref{fig:setup}(a). This effectively reduces the pulse
train period to $\tau/N$ with $\tau$ the cavity round-trip, while
circumventing the difficulties inherent in using shorter cavities.
These regular pulse arrangements can be identified spectrally since
the repetition rate multiplication corresponds to an equivalent $N$-fold
increase in the distance between spectral lines, that becomes a multiple
of the fundamental free spectral range (FSR), i.e. $N/\tau$. Such
states have been widely observed in mode-locked fiber lasers \citep{GG-JSB-97,HLS-OL-08,ASG-OL-11}
but also in optical systems with broken phase symmetry such as optically
injected Kerr micro-cavities \citep{CLD-NAP-17,KPG-NAP-19}. In the
latter, regular pulse trains, also termed ``soliton crystals'',
are phase locked to an external reference beam which leads to a unique,
well-defined, frequency comb. The phase invariance of the electromagnetic
field in a mode-locked laser radically modifies the picture as it
allows each pulse to possess a \textsl{different} phase $\varphi_{i}$
with $i\in\left[1,N\right]$, see Fig.~\ref{fig:setup}(b). 

In this letter, we ask the seemingly naive question: are the $N$
pulses circulating in a HML cavity necessarily coherent with each
other? What are their phase relations (if any) and, could an incoherent
mode-locked crystal of pulses simply exist?

\begin{figure*}
\includegraphics[width=1\textwidth]{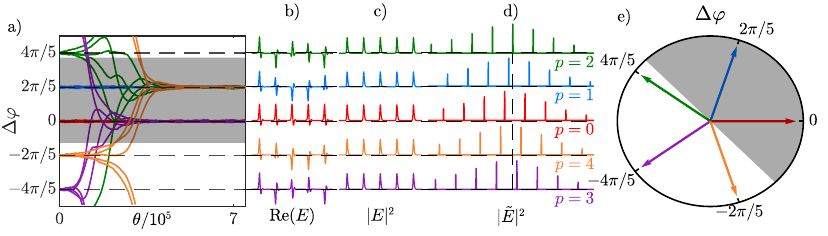}
\caption{a) Time evolution of various splay states associated to the HML$_{5}$
solutions obtained from solving Eqs.~(\ref{eq:Haus1}-\ref{eq:Haus2})
numerically. The initial phase differences are given by $\Delta\varphi_{p}=\frac{2\pi}{N}p,$
and are shown in panels b) and c) where one can see that the intensity
profiles are identical while the real parts of the electric fields
differ. d) Frequency spectrum of the initial conditions. For each
phase difference the corresponding comb is shifted by $\frac{p}{\tau}$.
e) Phase plane visualization of the initial conditions used in a)-d).
The shaded gray areas in panels a) and e) correspond to the region
of stability which is of size $\pi$. The trajectories in a) converge
to steady states within the region of stability. Parameters are $\left(\gamma_{f},k,\alpha,\beta,\tau,g_{0},\Gamma,q_{0},s,\sigma\right)=\left(40,0.8,1.5,0.5,12.5,4g_{\mathrm{th}},0.08,0.3,10,10^{-5}\right)$
and corresponds to a $5$~GHz mode-locked laser emitting $14\,$ps
pulses (FWHM).}
\label{fig:steady-states}
\end{figure*}

For the purpose of illustration, we shall consider a passively mode-locked
unidirectional ring laser as depicted in Fig.~\ref{fig:setup} and
in which a saturable absorber promote pulsed emission.{\small{} }The
multi-pulse HML dynamics can be studied using the well established
Haus master equation~\citep{haus00,PGG-NAC-20,Hausen20} that links
the evolution of the electric field $\left(E\right)$, the gain $\left(g\right)$
and absorber population $\left(q\right)$, {\small{}
\begin{align}
\partial_{\theta}E= & \left(\frac{1}{2\gamma_{f}^{2}}\partial_{z}^{2}+\frac{1-i\alpha_{g}}{2}g-\frac{1-i\alpha_{q}}{2}q-k\right)E+\sigma\xi,\label{eq:Haus1}\\
\partial_{z}g= & \Gamma\left(g_{0}-g\right)-g\left|E\right|^{2}\,,\,\partial_{z}q=q_{0}-q-sq\left|E\right|^{2}.\label{eq:Haus2}
\end{align}
}Here, $z$ and $\theta$ denote the fast and slow time scales which
describe the evolution within one round-trip and from one round-trip
to the next one, respectively. Further, the gain bandwidth is $\gamma_{f}$,
the cavity losses are $k$ and $\alpha_{g,q}$ correspond either to
the linewidth enhancement factors for semiconductor material or the
transition line detuning in atomic gain media. Time is normalized
to the absorber recovery time of $80\,$ps and the gain recovery rate
is denoted as $\Gamma$, while the ratio of the saturation energy
of the gain and the absorber media is $s$. For simplicity, we model
the spontaneous emission fluctuations and the mechanical vibrations
in the cavity potentially impacting coherence with additive white
Gaussian noise of the amplitude $\sigma$ in Eq.~(\ref{eq:Haus1}).
The equations (\ref{eq:Haus1},\ref{eq:Haus2}) were complemented
with the dynamical boundary conditions detailed in \citep{Hausen20}.
However, we note that identical conclusions were obtained using periodic
boundaries or by using instead the first principle time-delayed model
of \citep{VT-PRA-05}. The coherence of a regular HML state can be
measured by \textsl{the order parameter $b=\frac{1}{N}\sum_{j}\exp\left[i\left(\varphi_{j}-\varphi_{j-1}\right)\right],$}which
is equivalent to the first order field correlation $g^{\left(1\right)}\left(\frac{\tau}{N}\right)$
for a train of equidistant pulses that are identical up to a phase,
see Supplementary Material (SM). Perfect order corresponds to $\left|\left\langle b\right\rangle \right|=1$,
see Fig.~\ref{fig:setup}(c), whereas full disorder yields $\left|\left\langle b\right\rangle \right|=0$.
The bracket denotes a temporal average over many cavity round-trips.
In order to derive in which configurations the pulses can exist in
an HML regime, it is sufficient to consider the ring boundary conditions
in Fig.~\ref{fig:setup}(a). A steady state can only exists if the
phase difference $\Delta\varphi$ between neighboring pulses is constant.
This condition can only be fulfilled if $N\Delta\varphi=2\pi p$ with
$p\in\left[0,N-1\right]$. As such, the steady states are characterized
by the integer index $p$ or equivalently a phase difference $\Delta\varphi_{p}=\frac{2\pi}{N}p$,
which is the definition of \textsl{a splay state} \citep{BYM-CHA-21}.
We note that a phase shift $\Delta\varphi_{p}$ between pulses separated
by a distance $\Delta z=\tau/N$ corresponds to an offset of the carrier
frequency $\nu_{p}=\Delta\varphi_{p}/\left(2\pi\Delta z\right)=p/\tau$.
Consequently, these $N$ splay states correspond to frequency combs
that are shifted of $p$ times the fundamental FSR of the cavity,
see Fig.~\ref{fig:steady-states}(b-d).

Pulses interact via the active material dynamics. When a pulse crosses
over the amplifier, it depletes its available gain, that only partially
recovers before the arrival of the next pulse. This leads to a repulsion
between pulses \citep{KCB-JQE-98,JCM-PRL-16}. For short cavities
where $\Gamma\tau<1$, the positions of the pulses are tightly bound,
as in a crystal, to $z_{j}=j\tau/N$. In addition, pulses also possess
phase sensitive interactions with their neighbors via the overlap
of their decaying tails, see Fig.~\ref{fig:setup}(a). Assuming exponential
tails, this coherent effect scales as $\exp\left(-\tau_{s}/\tau_{p}\right)\ll1$
with $\tau_{s}$ the separation between pulses and $\tau_{p}$ the
pulse width. Consequently, the phase of each pulse evolves in relation
with that of its nearest neighbors over a time scale much larger than
the round-trip. We note that previous works demonstrated coherence
in a HML laser with a ratio $\tau_{s}/\tau_{p}\simeq23$ \citep{HLS-OL-08}
indicating that this nearest neighbor interaction is extremely weak.
While ratios up to $\tau_{s}/\tau_{p}\simeq50$ where reported in
\citep{GBG-JOSAB-03}, the relevance of this weak interaction must
be compared with the amount of spontaneous emission and technical
noise leading to decoherence. 

Intuition about the phase evolution can be obtained by projecting
the dynamics of the high-dimensional Haus Eqs. (\ref{eq:Haus1},\ref{eq:Haus2})
onto a low-dimensional subspace that correspond to the slow evolution
of the weakly coupled phase modes. This central manifold reduction
reads \begin{widetext}

\begin{eqnarray}
\partial_{\theta}\varphi_{j} & = & A_{+}\sin\left(\varphi_{j-1}-\varphi_{j}+\psi_{+}\right)+A_{-}\sin\left(\varphi_{j+1}-\varphi_{j}+\psi_{-}\right)+\sigma\xi_{j}\left(t\right),\quad\varphi_{0}=\varphi_{N},\:j\in\left[1,N\right].\label{eq:phase-model}
\end{eqnarray}

\end{widetext} We denote $\left(A_{\pm},\psi_{\pm}\right)$ the amplitudes
and phases of the coupling forces with the left and right neighboring
pulses. The parameters $\psi_{\pm}$ are inherited from the asymmetrical
and possibly chirped nature of the pulses. For $\psi_{\pm}\neq0$
the interactions are non-variational, i.e. Eq.~(\ref{eq:phase-model})
can not be derived from an interaction potential. In addition, since
the pulses are \emph{a priori} not symmetrical, $A_{+}\neq A_{-}$
which renders their interactions non-reciprocal. This situates our
work in the emerging field of non-conservative and non-reciprocal
interactions~\citep{FHL-NAT-21,K-NAN-23}. Finally, $\xi_{j}\left(t\right)$
is a Gaussian white noise whose amplitude $\sigma$ corresponds to
the projection of the original stochastic process in Eq.~(\ref{eq:Haus1})
over the phase dynamics. We note that Eq.~(\ref{eq:phase-model})
corresponds to the Kuramoto model with short range (nearest neighbor)
interactions, which is also known as the dissipative XY model, see
\citep{ABP-RMP-05} and reference therein.
\begin{figure}
\includegraphics[width=1\columnwidth]{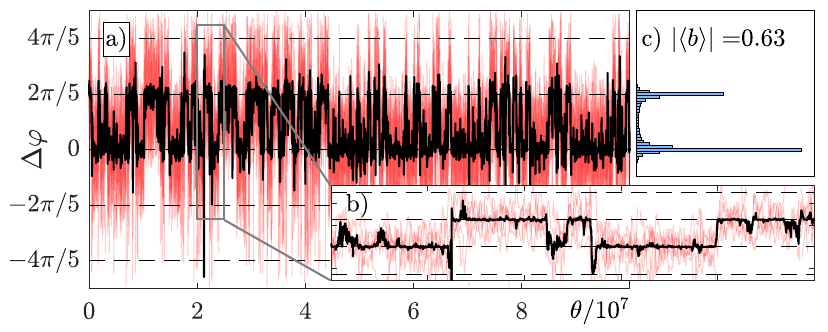}
\caption{a) Time evolution of phase differences of five pulses (light red)
for the same parameters as in Fig. \ref{fig:steady-states} obtained
from a direct numerical simulation of Eqs. (\ref{eq:Haus1},\ref{eq:Haus2})
with added noise of amplitude $\sigma=4\times10^{-4}$. Other parameters
are as in Fig.~1. The black line denotes the phase of the order parameter
$b$. As the system is bistable for $N=5$, the order parameter jumps
between the two stable steady states $\Delta\varphi_{0,1}=0,\frac{2\pi}{5}$.
b) Zoomed view around a region with multiple jumps. c) Statistical
distribution of the phases visited by $\arg b$. The two peaks correspond
to the two aforementioned steady states. }
\label{fig:bistability}
\end{figure}
 A linear stability analysis of Eq.~(\ref{eq:phase-model}) reveals
that a splay state is stable if $\alpha_{0}<\Delta\varphi_{p}<\alpha_{1}$
with
\begin{align}
\alpha_{k} & =k\pi+\arctan\left(\frac{A_{-}\cos\psi_{-}+A_{+}\cos\psi_{+}}{A_{-}\sin\psi_{-}-A_{+}\sin\psi_{+}}\right).\label{eq:stab1}
\end{align}
We observe that the range of stability is always $\pi$ and that the
half-circle of stability is simply rotated as a function of $\left(A_{\pm},\psi_{\pm}\right)$.
The most important consequence of Eq.~(\ref{eq:stab1}) is that the
splay states that are solutions of Eq.~(\ref{eq:phase-model}) can
be \emph{multistable}. In fact, for $N\ge5$, there must be at least
two states that fall within the range of stability and 50\% of states
are stable in the thermodynamic limit $N\rightarrow\infty$. In order
to test our predictions we performed numerical simulations of the
original Haus master Eqs.~(\ref{eq:Haus1},\ref{eq:Haus2}), which
we initialized with the five different splay states, see Fig.~\ref{fig:steady-states}(a,b).
We stress that all these solutions correspond to the same intensity
profile as demonstrated in Fig.~\ref{fig:steady-states}(c). The
rotation of the field $\Delta\varphi_{p}$ from one pulse to the next
creates a $p/\tau$ frequency-shifted comb, see Fig.~\ref{fig:steady-states}(d).
We observe that two initial conditions remain stable upon their time
evolution as they lie within the stable region defined by Eq.~(\ref{eq:stab1})
and marked by the gray areas in Fig.~\ref{fig:steady-states}(a,e).
The other three trajectories correspond to unstable states and converge
to one of the two stable configurations, cf. Fig.~\ref{fig:steady-states}(a).
The parameters $\left(A_{\pm},\psi_{\pm}\right)$ in Eq.~(\ref{eq:phase-model})
were extracted from Eqs.~(\ref{eq:Haus1},\ref{eq:Haus2}) by using
a perturbation analysis.

The prediction of multistability for $N\ge5$ has a profound consequence
on the measurement of the coherence as it must naturally be interpreted
by comparing the Kramers' escape time of each splay state \citep{G-BOOK-95}
and the duration of the measurement time. The latter typically occurs
experimentally over a time scale of a second, which corresponds to
billions of round-trips. We show in Fig.~\ref{fig:bistability}(a)
the result of a long simulation of Eqs.~(\ref{eq:Haus1},\ref{eq:Haus2})
over $10^{8}$ round-trips, i.e. $\sim0.4\,s$ for $\tau=4.4\,$ns.
Due to the presence of noise, the system is able to visit many times
all the stable states. This can be seen best in Fig.~\ref{fig:bistability}(b)
which provides a zoom around a small part of the original time trace.
We note that visiting the unstable states is also possible for higher
noise amplitudes and that the noise amplitude used here only give
rise to a $5~\%$ fluctuation of the pulse peak power (cf. Figs.~1,2
of the SM). We detail in Fig.~\ref{fig:bistability}(c) a histogram
of the value of $\arg b$. Due to the low noise value, the distribution
is narrowly peaked around each stable splay state and the theoretical
value of the order parameter should be $\left\langle b_{p}\right\rangle =\exp\left(i\Delta\varphi_{p}\right)$.
Indeed, if the value of the coherence $\left|\left\langle b\right\rangle \right|$
is calculated over a time smaller than the average residence time,
we found $\left|\left\langle b\right\rangle \right|\sim0.95$. If
instead the coherence is measured over the entire time trace, we obtain
$\left|\left\langle b\right\rangle \right|=0.63$ which is the result
of the partial cancellation between $\left\langle b_{0}\right\rangle $
and $\left\langle b_{1}\right\rangle $.

\begin{figure}
\includegraphics[width=1\columnwidth]{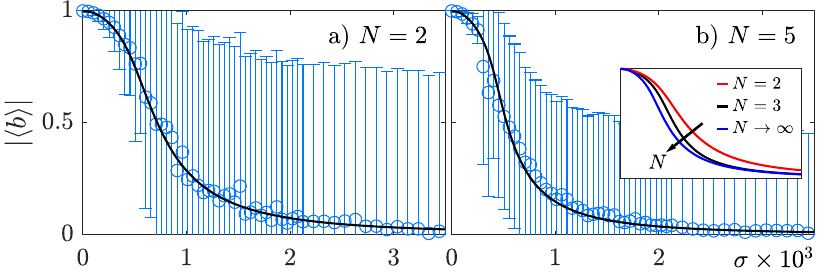}
\caption{The order parameter $\left|\langle b\rangle\right|$ as a function
of noise for different numbers of pulses in the cavity (blue: Haus
Eqs.~(\ref{eq:Haus1},\ref{eq:Haus2}), black: phase model in Eq.~(\ref{eq:phase-model})).
The cavity round-trip is increased according to the HML number as
$\tau=N\times2.5$. Same parameters as in Figs.~\ref{fig:steady-states}
and Fig.~\ref{fig:bistability}. Inset in b): analytical result for
the Hamiltonian phase model.}
\label{Fig:noise_b_scaling}
\end{figure}

\begin{figure*}[t]
\includegraphics[viewport=0bp 0bp 1712bp 487bp,width=1\textwidth]{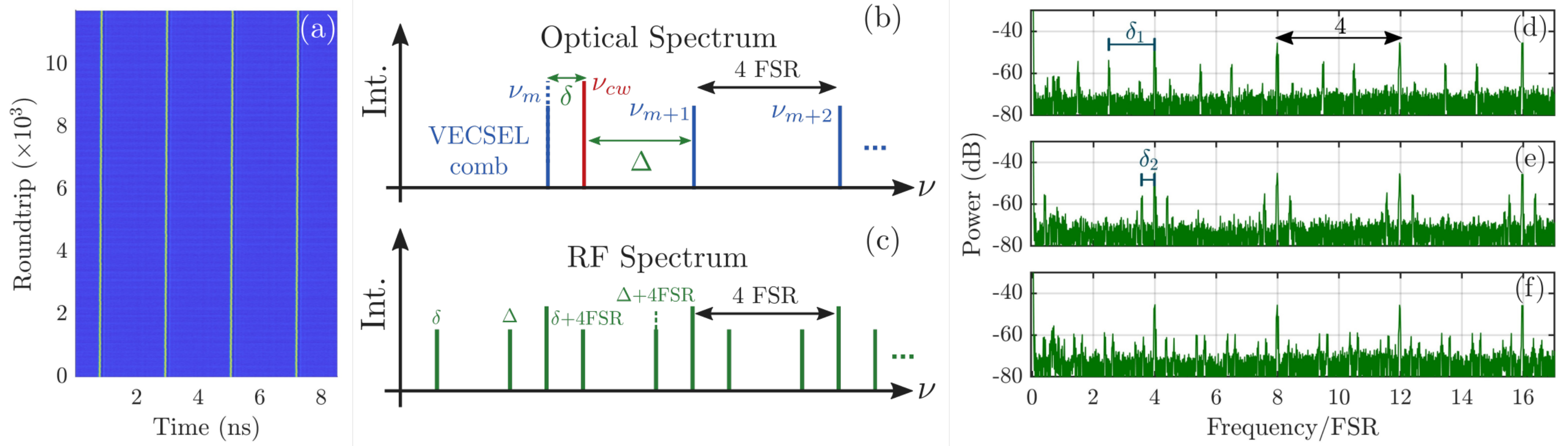}\caption{a) Space-time diagram of the laser output showing four equidistant
pulses. Central column. b) Spectral illustration of the heterodyne
beating between a splay state of four pulses and a CW laser. c) Illustration
of the resulting RF spectrum obtained from the heterodyne beating
in b). Right column: RF spectra of the heterodyne beating signal.
d,e) Two examples of coherent splay states with a spectral shift of
one FSR. f) Incoherent splay state characterized by the average spectral
emission composed by spectral lines separated by one FSR.}
\label{fig:exp}
\end{figure*}

Having demonstrated the good agreement between Eqs.~(\ref{eq:Haus1},\ref{eq:Haus2})
and Eq.~(\ref{eq:phase-model}) we used a combination of the two
to perform an extended analysis as a function of $N$ and the noise
level. As the number of stable states grows with $N$ bringing the
splay states closer together and diminishing the potential barrier
between them one can wonder how stable a single splay state can be
if, e.g. $N=350$ as discussed in \citep{HLS-OL-08}? Our results
are summarized in Fig.~\ref{Fig:noise_b_scaling}. We notice that
if the amplitude of the noise is too low, the system will not be able
to jump from one steady state to the other leading to a coherence
close to unity. On the other hand, if it is too high in comparison
with the force between pulses, i.e. $\sigma\gg A_{\pm}$, coherence
is entirely lost as $\left|\left\langle b\right\rangle \right|\rightarrow0$.
We note that in this case, the system continues to generate a very
regular intensity pulse train and it can therefore be interpreted
as \textsl{an incoherent crystal}. Finally, we confirm that the coherence
decays faster for larger values of $N$ as it is averaged over a larger
number of states. The numerical curves in Fig.~\ref{Fig:noise_b_scaling}
( in blue) were fitted with the analytical formula for the coherence
(in black) of the Hamiltonian XY model \citep{KT-JPC-73} that shares
the same steady states as Eq.~(\ref{eq:phase-model}) (cf. SM for
details).

To confirm our theoretical predictions, we implemented an experimental
setup consisting of a vertical external-cavity surface-emitting laser
mode-locked using a semiconductor saturable absorber mirror as in
\citep{BVM-Optica-22}. The laser cavity (see SM) has a round-trip
time $\tau=8.5\,$ns. For this value of $\tau$, the optical pulses
can be addressed on and off independently \citep{CSV-OL-18,BSV-OL-21}
although they become equidistant rapidly due to the repulsive incoherent
forces discussed above. This leads to the multistability of the HML
states with $N\in\left[0,8\right]$. The laser output is sent to a
heterodyne measurement setup that analyzes the spectral features of
the pulse train. By beating the laser output with a stable CW laser
source, we convert a portion of the optical spectrum into the RF domain
(see SM). We analyze the coherence of a pulse train consisting of
4 equidistant pulses. Figure \ref{fig:exp}(a) shows a spatio-temporal
map of the laser intensity for the HML$_{4}$ solution, where the
horizontal axis represents the round-trip time and the vertical axis
is the number of round-trips, which highlights that the pulses are
perfectly equispaced, have the same amplitude, and are stable over
more than 10 000 round-trips. Figure \ref{fig:exp}(b,c) describes
the principle of the heterodyne measurement. Panel (b) schematically
illustrates the optical spectrum of a mode-locked laser that is emitting
on a coherent splay state with $N=4$. Due to the coherent interaction
between the pulses, the spectrum is composed of equidistant lines
every 4 FSR. The CW laser frequency is represented in red and the
beating frequencies are shown in green. The corresponding RF signal
is shown in panel (c).

The jump from one coherent splay state $\Delta\varphi_{p}$ to another
$\Delta\varphi_{q}$ should manifest in the optical spectrum by a
$\left(p-q\right)$-harmonic FSR shift while the interference pattern
should keep the same visibility (see SM). The heterodyne optical spectra
(cf. the right column of Fig.~\ref{fig:exp}) reveal the various
splay states explored by the system. Figure~\ref{fig:exp}(d) shows
a situation identical to the example depicted in panel (c), which
indicates that the pulsed solution is a coherent splay state. In panel
(e), the RF spectrum shows the same features as panel (d) but the
lines in the optical spectrum are shifted with respect to the previous
case. In accordance with the theory, the shift experienced by the
optical lines corresponds to one FSR of the laser optical spectrum,
as demonstrated in the Supplementary Material. Figure \ref{fig:exp}
(f) reveals that the pulse train can also be incoherent, as predicted
from the theoretical analysis. This incoherent state is characterized
by phase jumps which are responsible for the emergence of multiple
beating lines in the optical spectrum. Here, the maximum number of
8 lines (2 beat tones times 4 possible FSR jumps) is obtained, which
indicates that within the acquisition time of the RF spectrum analyzer
(20~ms), the phase differences between the pulses have explored all
the four stable and unstable possible configurations (from 0 to $3\pi/4$),
i.e. the modal separation is one FSR. This observation demonstrates
that a pulse train can be stable in intensity, while the phase difference
between consecutive pulses can vary in time.

In conclusion, we have demonstrated that the harmonic pulse trains
emitted by a mode-locked semiconductor laser is equivalent to the
splay-phase states of the Kuramoto model with short range interactions.
The multistability between frequency combs was demonstrated experimentally
and we could observe the coherence parameter $\left|b\right|$ jumping
between high and vanishing values. This implies that the notion of
coherence must be interpreted by comparing the duration of the measurement
time with the Kramers' escape time of each splay state, which can
be extremely long due to the small amplitude of noise and of the weak
pulse interactions. We disclosed that this partially disordered state
for the phase of the optical field features regular train of pulses
in the field intensity, a state we termed an incoherent optical crystal.
Next, we showed how the laser cavity can experience a continuous shift
from coherence to incoherence as a function of the number of pulses
as well as the ratio between stochastic forces and the tail overlap
between pulses. The multistability induced by the splay phases can
potentially be controlled and used as a degree of freedom for encoding
information in mode-locked lasers. Finally, we stress that we used
a minimal model only considering nearest neighbor coupling which was
sufficient to show the transition from coherent to incoherent emission
and explain the experimental results. Additional couplings arising
from time-delayed feedback, as in \citep{BSV-OL-21}, or intra-cavity
lenses reflection would induce further non-local coupling between
pulses and lead to an even richer phenomenology, as observed in other
fields \citep{FKK-PRL-03,MPS-PRL-16,JMG-PRL-17}.
\begin{acknowledgments}
We acknowledge the financial support of the project KOGIT, Agence
Nationale de la Recherche (No. ANR-22-CE92-0009) and Deutsche Forschungsgemeinschaft
(DFG) via Grant No. 505936983. TGS and JJ acknowledge funding from
the Studienstiftung des Deutschen Volkes and the Ministerio de Economía
y Competitividad (PID2021-128910NB-100 AEI/FEDER UE, PGC2018-099637-B-100
AEI/FEDER UE), respectively.
\end{acknowledgments}


\end{document}


\title{Multistable Kuramoto splay states in a crystal of mode-locked laser pulses: Supplementary Material}
\author{T. G. Seidel}
\affiliation{Institute for Theoretical Physics, University of Münster, Wilhelm-Klemm-Str. 9,
	48149 Münster, Germany}
\affiliation{Departament de Física and IAC$^{3}$, Universitat de les Illes Balears,
	C/ Valldemossa km 7.5, 07122 Mallorca, Spain}
\author{A. Bartolo}
\affiliation{Institut d'Electronique et des Systèmes, CNRS UMR5214, 34000 Montpellier,
	France}
\author{A. Garnache}
\affiliation{Institut d'Electronique et des Systèmes, CNRS UMR5214, 34000 Montpellier,
	France}
\author{M. Giudici}
\affiliation{Université Côte d'Azur, CNRS, Institut de Physique de Nice, 06560
	Valbonne, France}
\author{M. Marconi}
\affiliation{Université Côte d'Azur, CNRS, Institut de Physique de Nice, 06560
	Valbonne, France}
\author{S. V. Gurevich}
\affiliation{Institute for Theoretical Physics, University of Münster, Wilhelm-Klemm-Str. 9,
	48149 Münster, Germany}
\author{J. Javaloyes}
\affiliation{Departament de Física and IAC$^{3}$, Universitat de les Illes Balears,
	C/ Valldemossa km 7.5, 07122 Mallorca, Spain}

\maketitle

\section{Motivation for the order parameter}

In order to measure the degree of order in a train of pulses, it is
instructive to introduce an order parameter. We consider for a configuration
of $N$ pulses the first order correlation function 
\begin{align}
g^{\left(1\right)}\left(r\right) & =\frac{\left\langle E^{*}\left(t\right)E\left(t+r\right)\right\rangle }{\left\langle \left|E\left(t\right)\right|^{2}\right\rangle }.
\end{align}
This function needs to be evaluated at the position of the neighboring pulse, i.e. when $N$
pulses are present in the cavity of size $\tau$, we consider $g^{\left(1\right)}\left(\frac{\tau}{N}\right)$.
Further, the average is defined as 
\begin{equation}
\left\langle E\left(t\right)\right\rangle =\frac{1}{\tau}\intop_{0}^{\tau}E\left(t\right)dt.
\end{equation}
Experimentally, this makes sense as the first order correlation function
can be obtained by interfering the electric field with itself after
a retarded time $\frac{\tau}{N}$. Next, we consider the electric field 
to consist of $N$ identical in shape and equidistant pulses. This
implies that there is no shape deformation mode. The equidistance
of the pulses implies that the translation mode is damped. This is
the case for pulses in a relatively short cavity, where pulses are
locked in position due to gain repulsion as it is the case in the
experiment presented in the main part of the paper. The remaining
degree of freedom is the phase $\varphi_{j}$ of the $j$-th pulse
such that the full field $E(t)$ reads

\begin{align}
	E\left(t\right)=\sum_{k=1}^{N}E_{0}\left(t-\left(k-1\right)\frac{\tau}{N}\right)e^{-i\varphi_{k}}\,,
\end{align}
where $E_{0}\left(t\right)$ is the optical field for a single pulse.
Further, we define $\bar{I}\equiv\left\langle \left|E_{0}\left(t\right)\right|^{2}\right\rangle $
as the average intensity of a single pulse. The first order correlation
function then reads 
\begin{align}
g^{\left(1\right)}\left(\frac{\tau}{N}\right) & =\frac{\left\langle \left[\sum_{k=1}^{N}E_{0}^{*}\left(t-\left(k-1\right)\frac{\tau}{N}\right)e^{i\varphi_{k}}\right]\left[\sum_{k=1}^{N}E_{0}\left(t-k\frac{\tau}{N}\right)e^{-i\varphi_{k}}\right]\right\rangle }{\left\langle \left|\sum_{k=1}^{N}E_{0}\left(t-k\frac{\tau}{N}\right)e^{i\varphi_{k}}\right|^{2}\right\rangle } \nonumber\\
 & =\frac{\left\langle \sum_{k=1}^{N}\left|E_{0}\left(t-k\frac{\tau}{N}\right)\right|^{2}e^{i\left(\varphi_{k+1}-\varphi_{k}\right)}\right\rangle }{N\bar{I}} \\
 & =\frac{\left\langle \left|E_{0}\left(t\right)\right|^{2}\right\rangle }{N\bar{I}}\sum_{k=1}^{N}e^{i\left(\varphi_{k+1}-\varphi_{k}\right)} \nonumber\\
 & =\frac{1}{N}\sum_{k=1}^{N}e^{i\left(\varphi_{k+1}-\varphi_{k}\right)}, \nonumber
\end{align}
where we used the periodic boundaries, i.e. $E_{0}\left(t\right)=E_{0}\left(t-\tau\right)$
and considered the overlap of the pulses to be small such that we
neglect all cross terms, i.e. $\left\langle \left|\sum_{k=1}^{N}E_{0}\left(t-k\frac{\tau}{N}\right)e^{i\varphi_{k}}\right|^{2}\right\rangle =N\bar{I}$.
Hence, the definition of the order parameter $b$ in the main text corresponds to $g^{\left(1\right)}\left(\frac{\tau}{N}\right)$:

\begin{eqnarray}
b & \equiv g^{\left(1\right)}\left(\frac{\tau}{N}\right)= & \frac{1}{N}\sum_{j=1}^{N}e^{i\left(\varphi_{j+1}-\varphi_{j}\right)}.
\end{eqnarray}
It makes sense to consider the limit cases of $b$. The modulus of
$b$ is 1, if all pulses have the same phase difference with respect
to their neighbor, i.e.
if $\varphi_{j+1}-\varphi_{j}=\beta=const.\ \forall j$,
then 
\begin{equation}
b=\frac{1}{N}\sum_{j=1}^{N}e^{i\left(\varphi_{j+1}-\varphi_{j}\right)}=e^{i\beta}.
\end{equation}
In contrast, the modulus of $b$ is 0 when the phase differences are
equally spread, i.e. $\varphi_{j+1}-\varphi_{j}=\frac{2\pi p}{N}j$ where
$p\in\mathbb{Z}$. Then
\begin{align}
b & =\frac{1}{N}\sum_{j=1}^{N}e^{i\frac{2\pi p}{N}j}=0.
\end{align}

\section{The order parameter as a function of the noise amplitude and the number of pulses}

In the presence of noise the order parameter $b$ is a statistical
quantity. However, we can still make predictions using the probability
distribution of the phases. In the variational reciprocal case, (i.e. $\psi_{\pm}=0$ and $A_{+}=A_{-}$, cf. Eq. (4) of the main text) which
reads
\begin{align}
\dot{\varphi}_{j} & =\sin\left(\varphi_{j-1}-\varphi_{j}\right)+\sin\left(\varphi_{j+1}-\varphi_{j}\right)+\sigma\xi_{j}\left(t\right) \nonumber\\
 & =F_{j}\left(\varphi_{1},\dots,\varphi_{N}\right)+\sigma\xi_{j}\left(t\right)
\end{align}
we can define a potential that satisfies $F_{j}=-\frac{\partial U}{\partial\varphi_{j}}$
\begin{align}
U & =-\int F_{j}\left(\varphi_{1},\dots,\varphi_{N}\right)d\varphi_{j} \nonumber\\
 & =-\int\left[\sin\left(\varphi_{j-1}-\varphi_{j}\right)+\sin\left(\varphi_{j+1}-\varphi_{j}\right)\right]d\varphi_{j}\\
 & =-\sum_{k=1}^{N}\cos\left(\varphi_{k-1}-\varphi_{k}\right). \nonumber
\end{align}
The Fokker--Planck equation for the probability density function
(PDF) of the phase distribution $P\left(\varphi_{1},\dots,\varphi_{N},t\right)$
reads 
\begin{eqnarray}
\frac{\partial P}{\partial t}+\sum_{j}\frac{\partial}{\partial\varphi_{j}}\left[F_{j}P\right] & = & \frac{\sigma^{2}}{2}\sum_{j}\frac{\partial^{2}P}{\partial\varphi_{j}^{2}}.
\end{eqnarray}
We solve for the steady state of the PDF, i.e. $\partial_{t}P=0$.
Introducing $\beta=\frac{2}{\sigma^{2}}$ and integrating a first
time leads to
\begin{eqnarray}
\sum_{j}F_{j}P & = & \frac{1}{\beta}\sum_{j}\frac{\partial P}{\partial\varphi_{j}}\\
\Leftrightarrow-\sum_{j}\frac{\partial U}{\partial\varphi_{j}}P & = & \frac{1}{\beta}\sum_{j}\frac{\partial P}{\partial\varphi_{j}}.
\end{eqnarray}
This equation has the solution
\begin{align}
P & =\mathcal{N}\exp\left(-\beta U\right)\,,
\end{align}
where $\mathcal{N}$ is a normalization constant. Its value is fixed
via the integral condition 
\begin{align}
1 & =\int P\left(\varphi_{1},\dots,\varphi_{N}\right)d\varphi_{1}\dots d\varphi_{N}
\end{align}
and therefore
\begin{equation}
\frac{1}{\mathcal{N}}=\int\exp\left(\beta\sum_{k}\cos\left(\varphi_{k-1}-\varphi_{k}\right)\right)d\varphi_{1}\dots d\varphi_{N}.
\end{equation}
To evaluate the integral we define $\theta_{k}=\varphi_{k-1}-\varphi_{k}$.
However, in this notation, one variable is dependent of the other
ones, i.e. we can write 
\begin{align}
\theta_{j} & =-\left(\varphi_{1}-\varphi_{N}\right)-\left(\varphi_{2}-\varphi_{1}\right)-\left(\varphi_{3}-\varphi_{2}\right)-\dots-\left(\varphi_{j-1}-\varphi_{j-2}\right)-\left(\varphi_{j+1}-\varphi_{j}\right)-\dots-\left(\varphi_{N}-\varphi_{N-1}\right) \nonumber\\
 & =-\sum_{k\ne j}\theta_{k}.
\end{align}
As a choice, we express $\theta_{N}$ in terms of the other phase
differences. Then we have
\begin{align}
\frac{1}{\mathcal{N}} & =2\pi\int\exp\left(\beta\sum_{k=1}^{N-1}\cos\theta_{k}+\beta\cos\left(\sum_{k=1}^{N-1}\theta_{k}\right)\right)d\theta_{1}\dots d\theta_{N-1} \nonumber \\
 & =2\pi\int\prod_{k=1}^{N-1}\exp\left(\beta\cos\theta_{k}\right)\exp\left(\beta\cos\left(\sum_{k=1}^{N-1}\theta_{k}\right)\right)d\theta_{1}\dots d\theta_{N-1}
\end{align}
Next, we use the modified Bessel function of first kind which are
defined as
\begin{equation}
I_{n}\left(x\right)=\frac{1}{\pi}\int_{0}^{\pi}\cos\left(n\theta\right)e^{x\cos\theta}d\theta.
\end{equation}
Further, we use the following expansion in terms of the Bessel functions
\begin{equation}
\exp\left(\beta\cos\theta\right)=\sum_{k=-\infty}^{\infty}I_{k}\left(\beta\right)e^{ik\theta}.
\end{equation}
With that we obtain
\begin{align}
	\begin{split}
		\frac{1}{\mathcal{N}}=&2\pi\int\left(\sum_{n=-\infty}^{\infty}I_{n}\left(\beta\right)\exp\left(in\theta_{1}\right)\right)\dots\left(\sum_{n=-\infty}^{\infty}I_{n}\left(\beta\right)\exp\left(in\theta_{N-1}\right)\right) \nonumber\\
		&\left(\sum_{n=-\infty}^{\infty}I_{n}\left(\beta\right)\exp\left(in\sum_{k=1}^{N-1}\theta_{k}\right)\right)d\theta_{1}\dots d\theta_{N-1}
	\end{split}\\
	\begin{split}
		=&2\pi\int\sum_{k_{1},k_{2},\dots,k_{N}=-\infty}^{\infty}I_{k_{1}}\left(\beta\right)I_{k_{2}}\left(\beta\right)\dots I_{k_{N}}\left(\beta\right)\exp\left(ik_{1}\theta_{1}\right)\exp\left(ik_{2}\theta_{2}\right)\dots\\
		&\exp\left(ik_{N-1}\theta_{N-1}\right)\exp\left(ik_{N}\sum_{l=1}^{N-1}\theta_{l}\right)d\theta_{1}\dots d\theta_{N-1}
	\end{split}\\
	\begin{split}
		=&2\pi\int\sum_{k_{1},k_{2},\dots,k_{N}=-\infty}^{\infty}I_{k_{1}}\left(\beta\right)I_{k_{2}}\left(\beta\right)\dots I_{k_{N}}\left(\beta\right) \nonumber \\
		&\exp\left[i\left(\left(k_{1}+k_{N}\right)\theta_{1}+\dots+\left(k_{N-1}+k_{N}\right)\theta_{N-1}\right)\right]d\theta_{1}\dots d\theta_{N-1}.
	\end{split}
\end{align}
This can also be expressed as the sum of $N-1$ integrals. Each of
these integrals has the form 
\begin{equation}
\int_{0}^{2\pi}e^{i\left(k+l\right)\theta}d\theta=2\pi\delta_{k,-l}.
\end{equation}
Finally, we obtain 
\begin{align}
\frac{1}{\mathcal{N}} & =\left(2\pi\right)^{N}\sum_{k_{1},k_{2},\dots,k_{N}=-\infty}^{\infty}I_{k_{1}}\left(\beta\right)I_{k_{2}}\left(\beta\right)\dots I_{k_{N}}\left(\beta\right)\delta_{k_{1},-k_{N}}\delta_{k_{2},-k_{N}}\dots\delta_{k_{N-1},-k_{N}} \nonumber \\
 & =\left(2\pi\right)^{N}\sum_{k_{N}=-\infty}^{\infty}\left(I_{-k_{N}}\left(\beta\right)\right)^{N-1}I_{k_{N}}\left(\beta\right)\\
 & =\left(2\pi\right)^{N}\sum_{k=-\infty}^{\infty}\left(I_{k}\left(\beta\right)\right)^{N} \nonumber
\end{align}
and with that the PDF reads:
\begin{align}
	P\left(\varphi_{1},\dots,\varphi_{N}\right)=\frac{1}{\left(2\pi\right)^{N}}\frac{\exp\left(\beta \sum_{k=1}^{N}\cos\left(\varphi_{k-1}-\varphi_{k}\right)\right)}{\sum_{k=-\infty}^{\infty}\left(I_{k}\left(\beta\right)\right)^{N}}.
\end{align}
Next, we want to compute the expectation value of the order parameter.
For that we can recycle the result for the normalization constant.
However, we need to define the derivative of the Bessel function first:
\begin{align}
\partial_{\beta}I_{k}\left(\beta\right) & =I_{k-1}\left(\beta\right)-\frac{k}{\beta}I_{k}\left(\beta\right) \nonumber \\
 & =\frac{k}{\beta}I_{k}\left(\beta\right)+I_{k+1}\left(\beta\right).
\end{align}
With this we obtain
\begin{align}
\left\langle b\right\rangle  & =\int_{0}^{2\pi}\left(\frac{1}{N}\sum_{k}e^{i\left(\varphi_{k-1}-\varphi_{k}\right)}\right)P\left(\varphi_{1},\dots,\varphi_{N}\right)d\varphi_{1}\dots d\varphi_{N} \nonumber \\
 & =\frac{\mathcal{N}}{N}\int_{-\pi}^{\pi}\left(\sum_{k}\cos\left(\varphi_{k-1}-\varphi_{k}\right)\right)\exp\left(\beta\sum_{k}\cos\left(\varphi_{k-1}-\varphi_{k}\right)\right)d\varphi_{1}\dots d\varphi_{N} \nonumber \\
 & =\frac{\mathcal{N}}{N}\int_{-\pi}^{\pi}\partial_{\beta}\exp\left(\beta\sum_{k}\cos\left(\varphi_{k-1}-\varphi_{k}\right)\right)d\varphi_{1}\dots d\varphi_{N} \nonumber\\
 & =\frac{\mathcal{N}}{N}\partial_{\beta}\int_{-\pi}^{\pi}\exp\left(\beta\sum_{k}\cos\left(\varphi_{k-1}-\varphi_{k}\right)\right)d\varphi_{1}\dots d\varphi_{N} \nonumber\\
 & =\frac{\mathcal{N}}{N}\partial_{\beta}\frac{1}{\mathcal{N}} \nonumber\\
 & =\left(2\pi\right)^{N}\frac{\mathcal{N}}{N}\partial_{\beta}\sum_{k=-\infty}^{\infty}\left(I_{k}\left(\beta\right)\right)^{N} \nonumber\\
 & =\left(2\pi\right)^{N}\mathcal{N}\sum_{k=-\infty}^{\infty}\left(I_{k}\left(\beta\right)\right)^{N-1}\partial_{\beta}I_{k}\left(\beta\right) \nonumber\\
 & =\left(2\pi\right)^{N}\mathcal{N}\sum_{k=-\infty}^{\infty}\left(I_{k}\left(\beta\right)\right)^{N-1}\left(I_{k-1}\left(\beta\right)-\frac{k}{\beta}I_{k}\left(\beta\right)\right) \nonumber\\
 & =\frac{\sum_{k=-\infty}^{\infty}\left(I_{k}\left(\beta\right)\right)^{N-1}\left(I_{k-1}\left(\beta\right)-\frac{k}{\beta}I_{k}\left(\beta\right)\right)}{\sum_{k=-\infty}^{\infty}\left(I_{k}\left(\beta\right)\right)^{N}}.
\end{align}
The last part of the sum vanishes because the sum is symmetric (from
$k=-\infty$ to $k=\infty$). So we remain with
\begin{equation}
\left\langle b\right\rangle =\frac{\sum_{k=-\infty}^{\infty}\left(I_{k}\left(\beta\right)\right)^{N-1}I_{k-1}\left(\beta\right)}{\sum_{k=-\infty}^{\infty}\left(I_{k}\left(\beta\right)\right)^{N}}.\label{eq:b_sigma}
\end{equation}
In the limit of $N\rightarrow\infty$, we find that the expression
converges to

\begin{align}
	\lim_{N\rightarrow\infty}\left\langle b\right\rangle =\frac{I_{1}\left(\beta\right)}{I_{0}\left(\beta\right)}.
\end{align}
Further, we can look at the special case $N=2$ where Eq.~(\ref{eq:b_sigma})
converges to 
\begin{align}	
	\lim_{N\rightarrow2}\left\langle b\right\rangle =\frac{I_{1}\left(2\beta\right)}{I_{0}\left(2\beta\right)}\,.
\end{align}

\subsection{Approximation for low noise}

An important limit case is the limit of a low noise level. For low noise
we have $\sigma\rightarrow0$ such that $\beta\rightarrow\infty$.
For large arguments the modified Bessel functions of first kind can
be expanded as
\begin{equation}
I_{k}\left(\beta\right)=\frac{e^{\beta}}{\sqrt{2\pi\beta}}\left[1-\frac{4k^{2}-1^{2}}{1\left(8\beta\right)}\left(1-\frac{4k^{2}-3^{2}}{2\left(8\beta\right)}\left(1-\frac{4k^{2}-5^{2}}{3\left(8\beta\right)}\left(1-\dots\right)\right)\right)\right].
\end{equation}
We consider terms up to $\sigma^{2}$, i.e. up to $\beta^{-1}$. Then
we remain with
\begin{align}
I_{k-1}\left(\beta\right) & =\frac{e^{\beta}}{\sqrt{2\pi\beta}}\left[1-\frac{4\left(k-1\right)^{2}-1}{8\beta}+\mathcal{O}\left(\beta^{-2}\right)\right] \nonumber \\
 & =\frac{e^{\beta}}{\sqrt{2\pi\beta}}\left[1-\frac{4k^{2}-1}{8\beta}-\frac{1-2k}{2\beta}+\mathcal{O}\left(\beta^{-2}\right)\right] \nonumber \\
 & =I_{k}\left(\beta\right)-\frac{e^{\beta}}{\sqrt{2\pi\beta}}\left(\frac{1-2k}{2\beta}+\mathcal{O}\left(\beta^{-2}\right)\right).
\end{align}
As we truncate all terms $\mathcal{O}\left(\beta^{-2}\right)$ we
can rewrite the second term in terms of a Bessel function
\begin{align}
\frac{1-2k}{2\beta}\frac{e^{\beta}}{\sqrt{2\pi\beta}} & +\mathcal{O}\left(\beta^{-\frac{5}{2}}\right)=\frac{1-2k}{2\beta}I_{k}\left(\beta\right)+\mathcal{O}\left(\beta^{-\frac{5}{2}}\right)
\end{align}
such that
\begin{equation}
I_{k-1}\left(\beta\right)=\left(1-\frac{1-2k}{2\beta}\right)I_{k}\left(\beta\right)+\mathcal{O}\left(\beta^{-\frac{5}{2}}\right).
\end{equation}
Then we have
\begin{align}
\left\langle b\right\rangle  & =\frac{\sum_{k=-\infty}^{\infty}\left(I_{k}\left(\beta\right)\right)^{N-1}\left(1-\frac{1-2k}{2\beta}\right)I_{k}\left(\beta\right)}{\sum_{k=-\infty}^{\infty}\left(I_{k}\left(\beta\right)\right)^{N}} \nonumber \\
 & =1-\frac{\sum_{k=-\infty}^{\infty}\left(I_{k}\left(\beta\right)\right)^{N}\frac{1-2k}{2\beta}}{\sum_{k=-\infty}^{\infty}\left(I_{k}\left(\beta\right)\right)^{N}}
\end{align}
Here, the term linear in $k$ vanishes as the sum is symmetric and
we remain with
\begin{align}
\left\langle b\right\rangle  & =1-\frac{1}{2\beta} \nonumber \\
 & =1-\frac{\sigma^{2}}{4}\,.
\end{align}

\section{Distribution of phases in the presence of high noise values}
\begin{figure}
    \centering
    \includegraphics[width=0.75\textwidth]{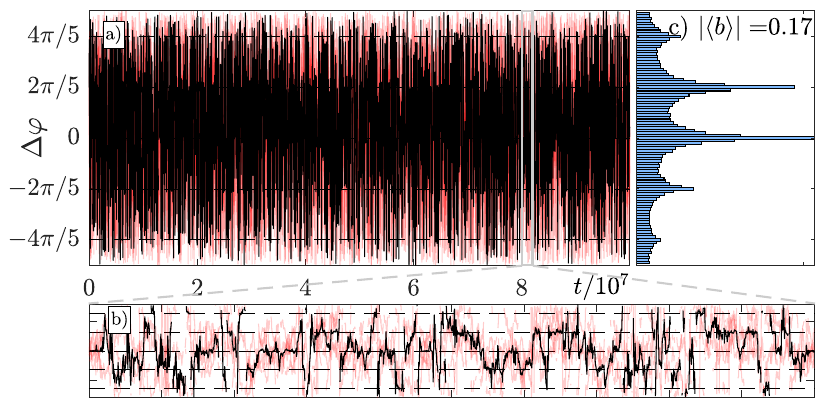}
    \caption{Time evolution of the Haus master equation with larger noise of the amplitude $\sigma=10^{-3}$ compared to figure 3 of the main text where $\sigma=4\times 10-4$.}
    \label{fig:bistability jump}
\end{figure}

\begin{figure}
    \centering
    \includegraphics[width=0.75\textwidth]{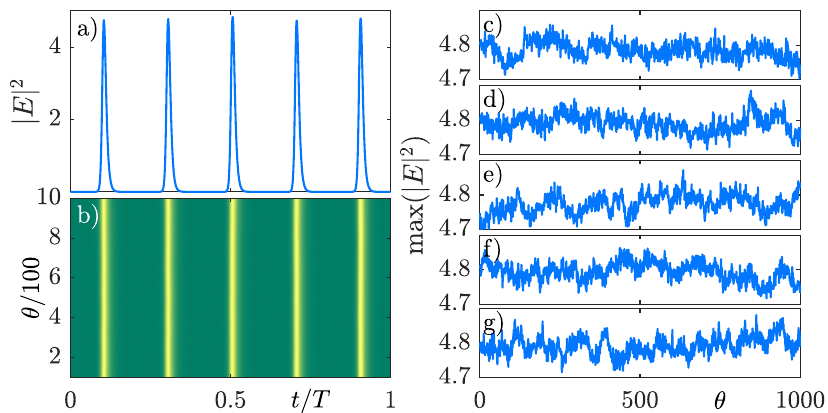}
    \caption{Short extract of the time trace from Fig. \ref{fig:bistability jump}. a) Profile at the last step of the simulation. b) Evolution of $\left|E\right|^2$ over $10^3$ roundtrips. c)-g) Evolution of the maxima of the five pulses in a),b).}
    \label{fig:bistability profile}
\end{figure}

Figure 3 of the main text shows the time evolution of the phase differences of five pulses in the Haus master equation with added noise of amplitude $\sigma=4\times 10-4$. The histogram in panel b) shows that the system mainly visits two phase configurations. In Fig. \ref{fig:bistability jump} we show the same plot for $\sigma=10^{-3}$. Here, the distribution broadens and also other states are visited. At the same time, the inset (panel b)) shows that the average residence time for the splay states becomes shorter compared to the corresponding figure in the main text. Even though Fig. \ref{fig:bistability jump} exhibits a very noisy behavior for the phases, the pulse train is still perfectly regular which is demonstrated in Fig. \ref{fig:bistability profile}. Here, an abstract over 1000 round-trips of of the simulation of Fig. \ref{fig:bistability jump} is shown in the intensity which exhibits fluctuations of $\approx 5\%$.

\section{Positions of the beat notes in the RF spectrum}

\begin{figure}
    \centering
    \includegraphics[width=0.75\textwidth]{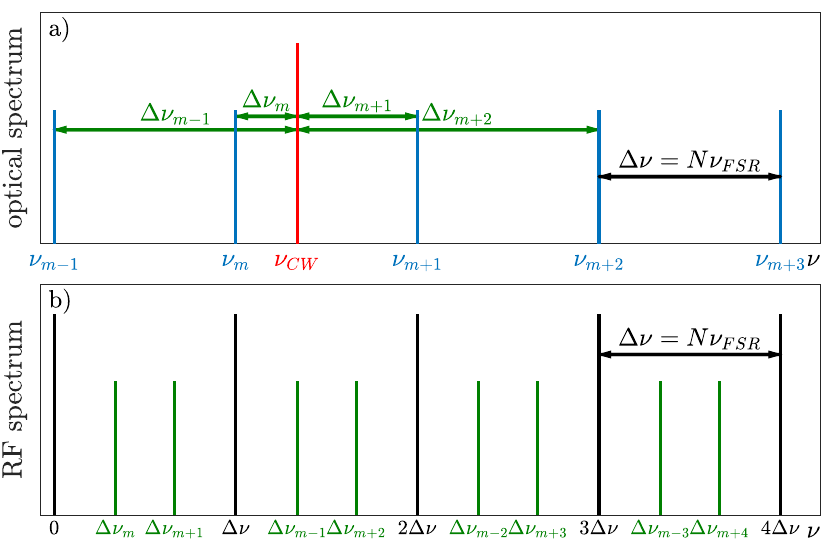}
    \caption{a) The optical spectrum of HML$_N$ state in a splay phase configuration where the peaks of the comb are separated by $N\nu_{FSR}$. We use a CW beam (red) to create beat notes (green). b) The RF spectrum with the beat notes shown in a).}
    \label{fig:sketch beating}
\end{figure}

For an HML$_{N}$ state in one of the splay phase configurations the spacing between neighboring teeth
in the spectrum is $N$ times the fundamental frequency $\nu_{FSR}=\frac{2\pi}{\tau}$:
$\Delta\nu=N\nu_{FSR}$. Therefore, the spectrum consists of peaks located
at $\nu_{k}=\nu_{0}+k\Delta\nu$, where $k\in\mathbb{Z}$ and $\nu_{0}$
is some offset around which the comb is centered. Further, in the
RF spectrum the main peaks are also separated by $\Delta\nu$ and
the CW beam has a frequency $\nu_{CW}$. Hence, the beat notes in
the RF spectrum for a given comb are found at $\Delta\nu_{k}=\left|\nu_{CW}-\nu_{k}\right|$.
In particular, the two beat notes with the lowest frequency (cf. the sketch in the main
text) are found for the value of $m$ such that $\nu_{m}<\nu_{CW}<\nu_{m+1}$
and hence, $\Delta\nu_{m}=\nu_{CW}-\nu_{0}-m\Delta\nu$ and $\Delta\nu_{m+1}=\nu_{0}-\nu_{CW}+\left(m+1\right)\Delta\nu$.
For simplicity, we define $\Delta\nu_{0}=\Delta\nu_{m}+m\Delta\nu=\nu_{CW}-\nu_{0}$.
Combining these results, we obtain 
\begin{equation}
\Delta\nu_{k}=\begin{cases}
+\Delta\nu_{0}-k\Delta\nu & k\le m\\
-\Delta\nu_{0}+k\Delta\nu & k>m
\end{cases}.
\end{equation}
This explains why in each interval of length $\Delta\nu$ in the RF
spectrum two beat notes are found corresponding to the interaction
with the teeth of the comb located to the left (index $m$) and to the right (index $m+1$) of
the CW frequency. For a visualization of the notation introduced above see Fig. \ref{fig:sketch beating}.

\subsection{Going from one frequency comb to the next one}

\begin{figure}
    \centering
    \includegraphics[width=0.75\textwidth]{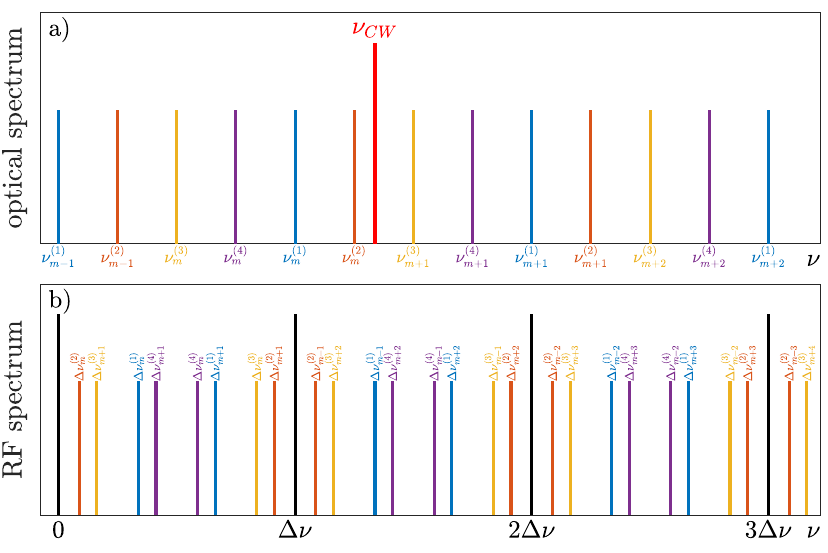}
    \caption{a) The optical spectrum of HML$_4$ state in a non-splay phase configuration. Here, the comb is a superposition of the four sub-combs shown in green, orange, yellow and purple.  The separation of two e.g. blue peaks is $\Delta\nu=4\nu_{FSR}$ while neighboring peaks (e.g. blue and orange) are separated by $\nu_{FSR}$. We use a CW beam (red) to create beat notes. b) The RF spectrum with the beat notes. In each interval of size $\Delta\nu$, we find 8 beat notes.}
    \label{fig:sketch beating jump}
\end{figure}

In the main text, we explained that each HML$_{N}$ solution has $N$
steady states corresponding to the phase differences $\Delta\varphi_{p}=\frac{2\pi p}{N}$
where $p=1,\dots,N$ which correspond to a shift of the frequency
comb by $p\nu_{FSR}=p\frac{2\pi}{\tau}$. Hence, $\nu_{0}$ is shifted by
$p\nu_{FSR}$. Therefore, we introduce the index $p$ in the previous
notation such that $\Delta\nu_{k}^{\left(p\right)}=\left|\nu_{CW}-\nu_k^{(p)}\right|$ where
$\nu_k^{(p)}=\nu_{0}-p\nu_{FSR}+k\Delta\nu$. Hence, in general, there are $2N$ possible positions for the beat notes in each interval of size
$\Delta\nu$ in the RF spectrum (cf. Fig. \ref{fig:sketch beating jump}).

Next, we want to make conclusions on the phase relation of the pulses
in two different RF spectra $p$ and $q$ based on the respective
locations of the beat notes. We consider the beat notes $\Delta\nu_{m,+}^{\left(p\right)}$
and $\Delta\nu_{m,+}^{\left(q\right)}$. As expected, the distance
between these quantities gives a multiple of $\Delta\nu$
\begin{align}
\Delta\nu_{m}^{\left(p\right)}-\Delta\nu_{m}^{\left(q\right)} & =\Delta\nu_{0}^{\left(p\right)}-m\Delta\nu-\Delta\nu_{0}^{\left(q\right)}+m\Delta\nu \nonumber \\
 & =\left(p-q\right)\nu_{FSR}
\end{align}
In the same way, we can also relate the left and the right beat notes
of two different spectra $p$ and $q$
\begin{align}
\Delta\nu_{m}^{\left(p\right)}+\Delta\nu_{m+1}^{\left(q\right)} & =\Delta\nu_{0}^{\left(p\right)}-m\Delta\nu-\Delta\nu_{0}^{\left(q\right)}+(m+1)\Delta\nu \nonumber \\
 & =\left(p-q\right)\nu_{FSR}+\Delta\nu.
\end{align}

\section{Experimental setup}

\begin{figure}
\centering
\includegraphics[height=4.5cm]{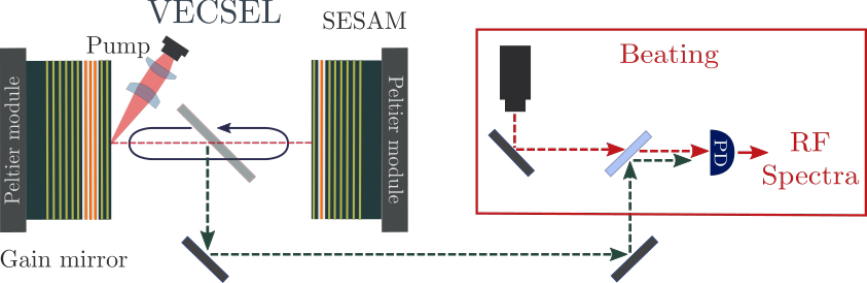}
\caption{Experimental setup of the mode-locked VECSEL and detection.  PEM: Piezo-Electric Mirror. PD: Photodiode. The heterodyne beating is realized between a tunable laser source (red arrow) and the VECSEL output (green arrow). }
\label{fig:exp}

\end{figure}

The experimental setup is shown in Fig. \ref{fig:exp}. It consists in a Vertical External-Cavity Surface-Emitting Laser (VECSEL) delimited by an optically pumped
InGaAs gain mirror emitting at 1060 nm and by a Semiconductor saturable Absorber Mirror (SESAM). The layout of the VECSEL cavity is similar to the one described in [1] and it is operated in the regime of temporal localized structures where multiple mode-locked pulses circulate in the resonator and each pulse can be individually addressed by shining short pump pulses inside the cavity [1]. For the heterodyne measurement, we use a tunable Littman-Metcalf diode laser system from Sacher with a linewidth smaller than 100 kHz.  The resolution
of the spectral detection is set by the RF power analyzer and amounts to 300 kHz.

\section{Experimental results with a splay state of 6 pulses}
\begin{figure*}
\centering
\includegraphics[height=6cm]{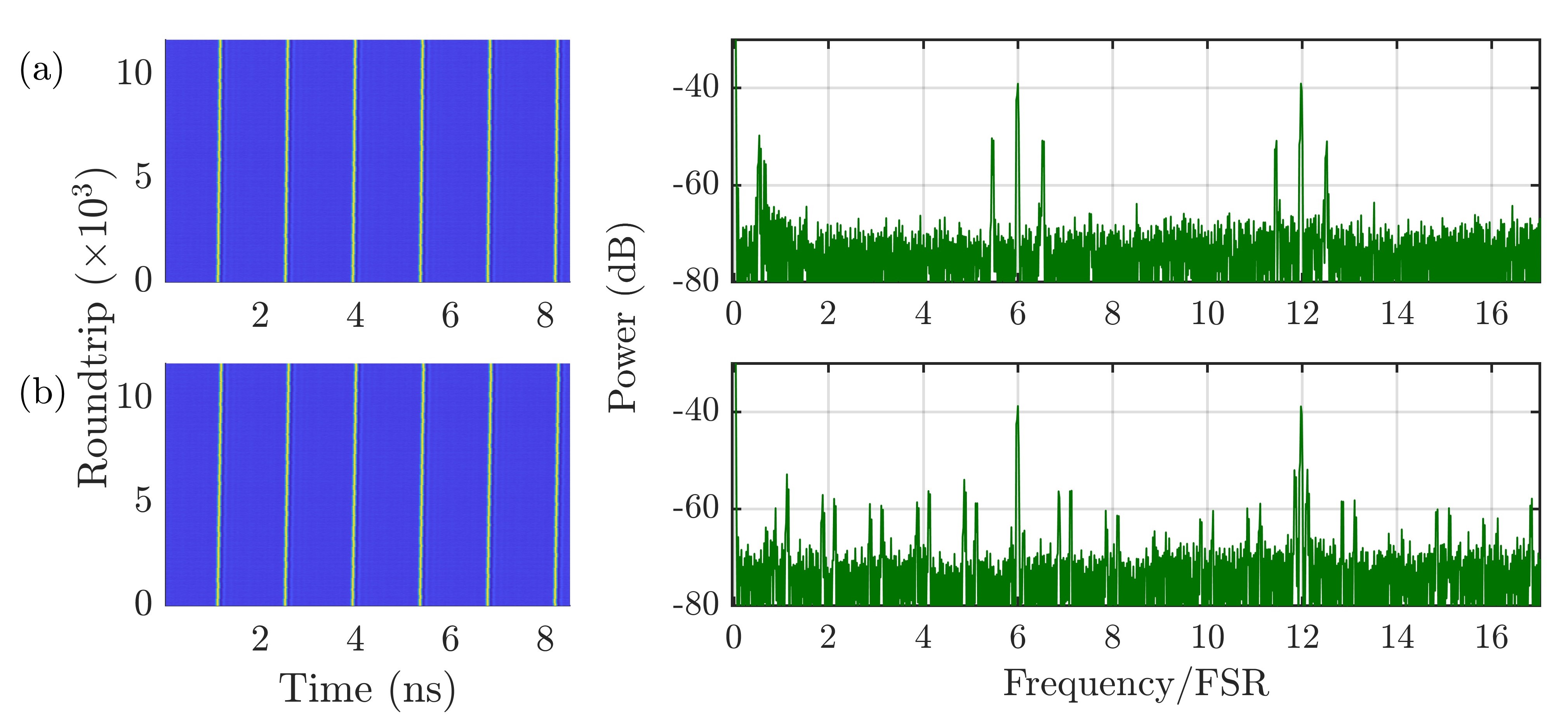}
\caption{Analysis of the splayed states of 6 pulses emitted by the mode-locked VECSEL. Left column: spatio-temporal maps of the laser intensity. Right column: RF spectrum of the heterodyne beating signal. 
a) Coherent splayed states. b) Incoherent state characterized by the average spectral emission composed by spectral lines separated by 1 FSR.}
\label{fig:6pulses}
\end{figure*}

Lastly, Fig.~\ref{fig:6pulses} shows that coherent splayed state and incoherent pulse trains can be obtained for a larger number of pulses in the external cavity. Similarly to the 4 pulses case, Fig.~\ref{fig:6pulses} shows a splayed state of 6 pulses (panel a) and an incoherent phase state (panel b).

\section{References}

[1] A. Bartolo, N. Vigne, M. Marconi, G. Beaudoin, K. Pantzas, I. Sagnes, G. Huyet, F. Maucher, SV Gurevich, J. Javaloyes, A. Garnache and M. Giudici, Temporal localized Turing patterns in mode-locked semiconductor lasers, Optica 9, 1386-1393 (2022).